\documentclass[acmsmall]{acmart}

\usepackage{url}

\AtBeginDocument{%
  \providecommand\BibTeX{{%
    \normalfont B\kern-0.5em{\scshape i\kern-0.25em b}\kern-0.8em\TeX}}}

\setcopyright{none}
\renewcommand\footnotetextcopyrightpermission[1]{} 
\begin{document}

\title[On Variants of Root Normalised Order-aware Divergence]{On Variants of Root Normalised Order-aware Divergence and a Divergence based on Kendall's Tau}

\author{Tetsuya Sakai}
\email{tetsuyasakai@acm.org}
\affiliation{%
  \institution{Waseda University}
  \streetaddress{3-4-1 Okubo, Shinjuku}
  \city{Tokyo}
  \country{Japan}
  \postcode{169-8555}
}




\begin{abstract}
This paper reports on a follow-up study of the work reported in \citet{sakai21acl,sakai21cikmlq},
which explored suitable evaluation measures for ordinal quantification tasks.
More specifically, the present study defines and evaluates, in addition to the quantification measures considered earlier,
a few variants of an ordinal quantification measure called
Root Normalised Order-aware Divergence (RNOD),
as well as a measure which we call Divergence based on Kendall's $\tau$ (DNKT).
The RNOD variants represent alternative design choices based on the idea
of Sakai's Distance-Weighted sum of squares (DW),
while DNKT
is designed to ensure 
that the system's estimated distribution over classes
is faithful to
the target priorities over classes.
As this Priority Preserving Property (PPP) of DNKT may be useful in some applications,
we also consider combining some of the existing quantification measures with DNKT.
Our experiments with eight ordinal quantification data sets
suggest that the variants of RNOD do not offer any benefit 
over the original RNOD at least in terms of system ranking consistency, 
i.e., robustness of the system ranking to the choice of test data.
Of all ordinal quantification measures considered in this study (including
Normalised Match Distance, a.k.a. Earth Mover's Distance),
RNOD is the most robust measure overall.
Hence the design choice of RNOD is a good one from this viewpoint.
Also, DNKT is the worst performer in terms of system ranking consistency.
Hence, if DNKT seems appropriate for a task,
sample size design should take its statistical instability into account.
\end{abstract}






\maketitle

\section{Introduction}\label{s:intro}

\emph{Quantification} (or \emph{prevalence estimation}) tasks~\cite{esuli10,gao16,sebastiani20}
require systems to estimate, for each test case with $N$ items, a gold distribution of the items over a given set of classes.
($N$ may vary across the test cases.)
When the classes are ordinal (e.g., \{\textit{negative}, \textit{neutral}, \textit{positive}\} or
\{1,2,3,4,5\} on a Likert scale),
the task is an \emph{ordinal quantification} task.
Examples of ordinal quantification tasks
include the SemEval 2017 Task~4 Subtask~E~\cite{SemEval:2017:task4},
the Dialogue Breakdown Detection Challenge~\cite{YuikoTsunomori2020DSI-G},
and the NTCIR Dialogue Evaluation task~\cite{zeng20}.
\emph{Nominal quantification} measures
such 
as the 
\emph{Kullback-Leibler Divergence} (KLD),
\emph{Jensen-Shannon Divergence} (JSD),
\emph{Root Normalised Sum of Squares} (RNSS, which is essentially \emph{Root Mean Squared Error}), and
\emph{Normalised Variational Distance} (NVD, which is essentially \emph{Mean Absolute Error})
are clearly not suitable for ordinal quantification tasks~\cite{sakai18sigir}:
note, for example, that 
these measures cannot consider the fact that
misinterpreting a 
\textit{negative} instance as a \textit{positive} instance is more serious than
misinterpreting it as a \textit{neutral} instance.

This paper reports on a follow-up study of the work reported in \citet{sakai21acl,sakai21cikmlq},
which explored suitable evaluation measures for ordinal quantification tasks.
More specifically, the present study defines and evaluates, in addition to the quantification measures considered earlier,
a few variants of an ordinal quantification measure called
\emph{Root Normalised Order-aware Divergence} (RNOD),
as well as a measure which we call \emph{Divergence based on Kendall's $\tau$} (DNKT).
The RNOD variants represent alternative design choices based on the idea
of Sakai's \emph{Distance-Weighted sum of squares} (DW),
while DNKT
is designed to ensure 
that the system's estimated distribution over classes
is faithful to
the target priorities over classes.
As this property of DNKT may be useful in some applications,
we also consider combining some of the existing quantification measures with DNKT.
Our experiments with eight ordinal quantification data sets
suggest that the variants of RNOD do not offer any benefit 
over the original RNOD in terms of system ranking consistency, 
i.e., robustness of the system ranking to the choice of test data~\cite{sakai21ecir}.
Of all ordinal quantification measures considered in this study (including
\emph{Normalised Match Distance} (NMD)~\cite{sakai18sigir}, a.k.a. \emph{Earth Mover's Distance}),
RNOD is the most robust measure overall.
Hence the design choice of RNOD is a good one from this viewpoint.
Also, DNKT is the worst performer in terms of system ranking consistency.


\section{Quantification Measures}\label{s:measures}

Table~\ref{t:quantification-measures} provides an overview of the properties of the quantification measures considered in the present study.
Note that a ``$\checkmark$'' does not necessarily mean an advantage:
for example, symmetry is not a required property for quantification tasks,
since we know which of the two distributions is the gold one in these tasks.
Section~\ref{ss:ordinal-quantification-measures} 
defines ordinal quantification measures (the first five measures in Table~\ref{t:quantification-measures}, including RNOD2, RNADW, RNADW2 which are evaluated for the first time in this paper);
Section~\ref{ss:nominal-quantification-measures}  defines
nominal quantification measures (NVD, RNSS, and JSD);
Section~\ref{ss:DNKT} introduces DNKT, a divergence measure based on Kendall's $\tau$,
which only considers the priorities across classes
and is not strictly a quantification measure.

\begin{table}[t]
\centering
\caption{Properties of the measures considered in this study:\\
(a)~considers the magnitude of the error in each class;\\
(b)~suitable for ordinal classes;\\
(c)~is symmetric;\\
(d)~does not assume equidistance for ordinal classes;\\
``N/A'' means ``not applicable.''
}\label{t:quantification-measures}\vspace*{-2mm}
\begin{tabular}{l|ccccc|ccc|c}
\toprule
	&NMD		&RNOD		&RNOD2		&RNADW		&RNADW2	&NVD		&RNSS		&JSD		&DNKT\\
\hline
(a)	&\checkmark	&\checkmark	&\checkmark	&\checkmark	&\checkmark	&\checkmark	&\checkmark	&\checkmark	&\\
(b)	&\checkmark	&\checkmark	&\checkmark	&\checkmark	&\checkmark	&			&			&			&\\
(c)	&\checkmark	&			&			&\checkmark	&\checkmark	&\checkmark	&\checkmark	&\checkmark	&\checkmark\\
(d)	&\checkmark	&			&\checkmark	&			&\checkmark	&N/A		&N/A		&N/A	&N/A\\
\bottomrule
\end{tabular}
\end{table}

\subsection{Ordinal Quantification Measures}\label{ss:ordinal-quantification-measures}

Let $C$ denote a set of ordinal classes,
represented by consecutive integers for convenience.
Let $p_{i}$ denote the estimated probability for Class~$i$, so that $\sum_{i \in C}p_{i}=1$.
Similarly, let $p_{i}^{\ast}$ denote the gold probability.
We also denote the entire probability mass functions by $p$ and $p^{\ast}$, respectively.
Let ${\it cp}_{i}=\sum_{k \leq i} p_{k}$, and
${\it cp}_{i}^{\ast}=\sum_{k \leq i} p_{k}^{\ast}$.
NMD
is given by~\cite{sakai18sigir}:
\begin{equation}\label{eq:nmd}
{\it NMD}(p, p^{\ast}) = \frac{
\sum_{i \in C} | {\it cp}_{i} - {\it cp}_{i}^{\ast} |
}
{
|C|-1
} \ .
\end{equation} 
NMD is a normalised form of Earth Mover's Distance, also known as 
\emph{Wasserstein} or \emph{Mallows Distance}~\cite{levina01,werman85}.

RNOD~\cite{sakai21acl,sakai21cikmlq} is defined as follows.
First, let the \emph{Distance-Weighted sum of squares} for Class~$i$ be:
\begin{equation}\label{eq:DW}
{\it DW}_{i} = \sum_{j \in C} \delta_{ij} (p_{j}-p_{j}^{\ast})^2 \ , 
\end{equation}
where
\begin{equation}\label{eq:equidistance}
\delta_{ij}=|i-j| \ .
\end{equation}
$\textit{DW}_{i}$ was designed 
to quantify the overall error
from the viewpoint of a particular gold class~$i$:
it tries to measure 
how much of its probability $p^{\ast}_{i}$
has been misallocated to other classes $j \in C (j \neq i)$,
by
assuming 
that the difference between
$p_{j}$ and $p_{j}^{\ast}$ is directly
caused by a misallocation of part of  $p^{\ast}_{i}$;
the weight $\delta_{ij}$ 
is designed to penalise the misallocation 
based on the distance between the ordinal classes.

Let $C^{\ast} = \{ i \in C | p_{i}^{\ast}>0\}$. 
That is,
$C^{\ast} (\subseteq C)$ is the set of classes with a non-zero gold probability.
\emph{Order-aware Divergence} is defined as:
\begin{equation}\label{eq:OD}
{\it OD}(p \parallel p^{\ast}) = \frac{1}{|C^{\ast}|} \sum_{i \in C^{\ast}} {\it DW}_{i} \ .
\end{equation}
Note that this is the average of $\textit{DW}_{i}$ over $C^{\ast}$ rather than over $C$ because,
as mentioned earlier, 
$\textit{DW}_{i}$ was designed 
to quantify the overall error
from the viewpoint of a particular \emph{gold} class~$i$.
RNOD is then defined as:
\begin{equation}\label{eq:RNOD}
{\it RNOD}(p \parallel p^{\ast}) = \sqrt{ 
\frac{
{\it OD}(p \parallel p^{\ast})
}
{
|C|-1
}
} \ .
\end{equation}
Although a symmetric version of RNOD called RSNOD is available
(based on symmetric order-aware divergence given by $\textit{SOD}(p, p^{\ast})=(\textit{OD}(p \parallel p^{\ast})+\textit{OD}(p^{\ast} \parallel p))/2$)~\cite{sakai18sigir},
we do not consider RSNOD in our experiments because
(a)~symmetry is not required for quantification evaluation; and
(b)~\citet{sakai21acl,sakai21cikmlq} showed that introducing symmetry is not beneficial in terms of system ranking consistency (i.e., robustness of the system rankings to the choice of test data)~\cite{sakai21ecir}.

Eq.~\ref{eq:equidistance} assumes equidistance.
If one prefers to avoid this assumption (since ordinal classes may not be interval classes),
the following is a natural alternative~\cite{sakai21cikmlq}.
\begin{equation}\label{eq:non-equidistance}
\delta_{ij} = \left( \sum_{k=\min(i,j)}^{\max(i,j)} p^{\ast}_{k} \right) - \frac{p^{\ast}_{i} + p^{\ast}_{j}}{2} \ .
\end{equation}
That is, we utilise the gold propabilities that lie between Classes $i$ and $j$
to define the distance.
This resembles the distance function used in Krippendorff's alpha for ordinal classes~\cite{krippendorff18}.
If Eq.~\ref{eq:non-equidistance} is used instead of Eq.~\ref{eq:equidistance}
when computing Eq.~\ref{eq:RNOD},
we call the resultant measure \emph{RNOD2} for convenience.

\citet{sakai21cikmlq} defined yet another variant of RNOD based on simply averaging 
the $\textit{DW}$'s over the entire set of classes $C$, instead of over $C^{\ast}$ (\textit{cf.} Eq.~\ref{eq:OD}).
\footnote{
This was a suggestion from one of the reviewers of the CIKM 2021 Learning to Quantify (LQ) workshop.
Independently, one of my students also asked me why OD is not defined as the average of $\textit{DW}_{i}$
over the whole set of classes $C$. But my original intention was to define $\textit{DW}_{i}$ for
each \emph{gold} class ($\in C^{\ast}$), for the reason discussed earlier.
}
\begin{equation}\label{eq:ADW}
{\it ADW}(p , p^{\ast}) = \frac{1}{|C|} \sum_{i \in C} {\it DW}_{i} \ ,
\end{equation}
From Eqs.~\ref{eq:DW} and~\ref{eq:ADW}, it is clear that ADW (``Average DW'') is symmetric.\footnote{
Similarly, it is clear from Eqs.~\ref{eq:DW} and~\ref{eq:OD} that
$C^{\ast}=C$ (i.e., there is no gold probability that is zero) is a sufficient 
condition for OD to be symmetric~\cite{sakai17evia-dialogues}.
Another sufficient condition for guaranteeing the symmetry of 
OD is: $|C^{\ast}|=1$ \emph{and} $|\{ p_{i} \in C \mid p_{i} >0\}|=1$ (i.e., 
both the gold and estimated distributions have exactly one positive probability).
} Root Normalised ADW (RNADW) is then defined as:
\begin{equation}\label{eq:RNADW}
\textit{RNADW}(p, p^{\ast}) = \sqrt{\frac{\textit{ADW}(p, p^{\ast})}{|C|-1}} \ .
\end{equation}
Furthermore, we also consider replacing Eq.~\ref{eq:equidistance} with Eq.~\ref{eq:non-equidistance} 
when computing the above, and call the resultant measure \emph{RNADW2}.
That is, like RNOD2, RNADW2 avoids the equidistance assumption.

\subsection{Nominal Quantification Measures}\label{ss:nominal-quantification-measures}

The three measures defined below ignore the ordinal nature of the classes~\cite{sakai18sigir}
and therefore should not be used for ordinal quantification tasks.
However, including them in our experiments is useful
for us to highlight the different properties of the ordinal quantification measures discussed earlier.
For example, while the system ranking according to NMD is very different from
those according to these nominal quantification measures,
the system ranking according to RNOD is less so~\cite{sakai21acl}.
Moreover, when we examine the system ranking consistency (i.e., robustness to the choice of test data)
of the ordinal quantification measures,
it is informative to see how they perform relative to nominal quantification measures.

\emph{Normalised Variational Distance} (NVD)~\cite{sakai18sigir} is essentially the Mean Absolute Error (MAE):
\begin{equation}\label{eq:nvd}
{\it NVD}(p, p^{\ast}) = \frac{1}{2} \sum_{i \in C} | p_{i} - p_{i}^{\ast}| \ .
\end{equation}
\emph{Root Normalised Sum of Squares} (RNSS)~\cite{sakai18sigir} is essentially the Root Mean Squared Error (RMSE):
\begin{equation}
{\it RNSS}(p, p^{\ast}) = \sqrt{ \frac{
\sum_{i \in C} (p_{i} - p_{i}^{\ast})^2
}
{
2
}
} \ .
\end{equation}
The advantages of RMSE over MAE is discussed in \citet{chai14}.

The \emph{Kullback-Leibler divergence} (KLD) for system and gold probability distributions over classes is given by:
\begin{equation}
{\it KLD}( p \parallel p^{\ast} ) = \sum_{i \in C \,\,\,\mathrm{s.t.}\,\,\, p_{i}>0} p_{i} \log_{2} \frac{p_{i}}{p_{i}^{\ast}} \ .
\end{equation}
As this is undefined if $p_{i}^{\ast}=0$, 
we only consider the 
more convenient
\emph{Jensen-Shannon divergence} (JSD)~\cite{lin91}:
\begin{equation}
{\it JSD}(p, p^{\ast}) = \frac{
{\it KLD}(p \parallel p^{M}) + {\it KLD}(p^{\ast} \parallel p^{M})
}
{
2
} \ ,
\end{equation}
where $p_{i}^{M}=(p_{i}+p_{i}^{\ast})/2$. Note that, unlike KLD, JSD is symmetric and bounded.

\subsection{DNKT: Divergence based on Kendall's $\tau$}\label{ss:DNKT}

Finally, consider a divergence measure that is not strictly a quantification measure,
as it does not consider the quantity in each class:
it only cares about the priorities across classes.
While such a measure is not adequate for ordinal quantification,
it may be useful in some other contexts, for example, in \emph{group-fair ranking evaluation}~\cite{sakai22arxiv-groupfairness}.
We decided to examine the property of this measure along with the ordinal and nominal quantification measures
described earlier.

This measure
 is based on Kendall's $\tau\mbox{-}b$ rank correlation~\cite{kendall62} and we therefore call it
 DNKT (Divergence based on Kendall's $\tau$).
We sort the bins of the gold distribution by the probabilities; we do the same with the estimated distribution;
and we finally compute the rank correlation between the two sorted lists, based on the number of concordant pairs of bins ($\textit{CONC}$)
and the number of discordant pairs of bins ($\textit{DISC}$) across the two lists. Note that,  when counting $\textit{CONC}$ and $\textit{DISC}$,
any bin pair is ignored if it is tied in at least one of the lists.
We use $\tau\mbox{-}b$ rather than the regular $\tau$ in order to handle tied bin pairs into account, for this is a required feature
for quantification evaluation. For example, consider a gold uniform distribution: all pairs of bins are tied so $\textit{CONC}=\textit{DISC}=0$ by definition.
While $\tau\mbox{-}b$ has the range of $[-1, 1]$, we transform it into a divergence measure where 
zero means perfect match.\footnote{
But, as we shall discuss later, 
DNKT is not zero even if both distributions are uniform.
}


Formally,
let $\textit{TIED}$ and $\textit{TIED}^{\ast}$
denote the number of tied pairs of bins in the sorted lists of estimated and gold probabilities, respectively.
Then the number of  bin pairs that are \emph{not} tied in each list is given by:
\begin{equation}
\textit{notTIED} = 
\frac{|C|(|C|-1)}{2} - \textit{TIED} \ , \,\,\,\,\, \textit{notTIED}^{\ast} = \frac{|C|(|C|-1)}{2} - \textit{TIED}^{\ast} \ .
\end{equation}
We define $\tau\mbox{-}b$ as follows.
\begin{equation}\label{eq:tau-b}
\tau\mbox{-}b = \frac{
\textit{CONC} - \textit{DISC}
}
{
\sqrt{\max(1, \textit{notTIED} )}
\sqrt{\max(1,  \textit{notTIED}^{\ast})}
} \ .
\end{equation}
Note that we have introduced the $\max$ operators to avoid division by zero:
recall that a uniform distribution says that every pair of bins is tied.

A normalised version of $\tau\mbox{-}b$ with a $[0,1]$ range can be obtained as:
\begin{equation}\label{eq:NKT}
\textit{NKT} = \frac{\tau\mbox{-}b + 1}{2} \ ,
\end{equation}
and a Divergence based on NKT (DNKT) can be defined as:
\begin{equation}\label{eq:DNKT}
\textit{DNKT} = 1 - \textit{NKT} = 
\frac{
1 - \tau\mbox{-}b
}
{
2
} \ .
\end{equation}

DNKT completely ignores the magnitude of the error for each bin,
unlike any other measure discussed above.
For example, if the gold distribution is $(0.4, 0.3, 0.2, 0.1)$
and the estimated distribution is $(0.31, 0.30, 0.20, 0.19)$, 
the estimated distribution is considered perfect.
As was mentioned earlier, we thought that 
this divergence may be useful in some contexts other than quantification tasks,
such as group-fair ranking tasks.
For example, if the target distribution for ranking scholarship applicants~\cite{sakai22arxiv-groupfairness}
says that the low-income group should be prioritised over middle-income group,
which in turn should be prioritised over high-income groups,
DNKT only cares about these priorities.

Because 
the above \emph{Priority Preserving Property} (PPP) of DNKT may be desirable in some contexts
and yet not satisfied by any of the quantification measures,
we also explored the idea of
combining DNKT with a quantification measure 
so that the hybrid measure inherits properties of both component measures.
More specifically, we combine Measure $M$ with DNKT as their harmonic mean
if at least one of the measures is non-zero:
\begin{equation}\label{eq:harmonic-mean}
\textit{DNKT\_}M = 
\frac{
2 * \textit{DNKT} * M
}
{
\textit{DNKT} + M
} \ .
\end{equation}
If $\textit{DNKT}=M=0$, we let $\textit{DNKT\_}M =0.$
For $M$, we consider JSD (representing a nominal quantification measure), NMD, and RNOD (ordinal quantification measures).

It should be noted, however, that DNKT is not useful if the gold distribution is uniform: in general, when at least one of the two distributions is uniform, $\textit{CONC} = \textit{DISC}=0$ and hence $\tau\mbox{-}b=0$ and $\textit{DNKT}=0.5$.

\section{Data}

\begin{table*}[t]
\centering
\caption{Eight data sets used in our experiments (C: Chinese; E: English; C+E: runs for both languages combined).}\label{t:taskdata}

\begin{small}
\begin{tabular}{lllccrrl}
\toprule
Short name	&Evaluation		&Task (Subtask)			&Task	&Language	&\#Ordinal	&Test data		&\#Runs used\\
in this paper	&venue			&					&Type	&		&classes	&sample size	&\\
\midrule
Sem16T4E	&SemEval-2016	&Task 4 		&OQ		&E		&5		&100			&12 \\
			&				&(Subtask E)	&			&		&		&				&\\
\midrule
Sem17T4E	&SemEval-2017	&Task 4		&OQ		&E		&5		&125			&14 \\
			&				& (Subtask E)	&			&		&		&				&\\
\midrule
STC-3 		&NTCIR-14	&Short Text  		&OQ		&C$+$E		&5		&390	&19 (10+9)\\
			&			&Conversation 3	&			&			&		&		&\\
DQ-\{A, E, S\}& (2019)				&(Dialogue 	&			&			&		&			&\\
			&					&Quality)		&			&			&		&			&\\
\midrule
DialEval-1	&NTCIR-15	&Dialogue 		&OQ		&C$+$E		&5		&300		&22 (13+9)\\
DQ-\{A, E, S\}&(2020)				&Evaluation 1		&			&			&		&			&\\
 			& 		&(Dialogue 			&			&			&		&			&\\	
 			&					&Quality)				&			&			&		&			&\\
\bottomrule
\end{tabular}
\end{small}

\end{table*}

The ordinal quantification data sets we use in our experiments are the same as those used in \citet{sakai21cikmlq};
they are briefly described in Table~\ref{t:taskdata}.
As can be seen, all data sets come with five ordinal classes.
For the two SemEval data sets,
the classes are tweet polarities, namely,
\emph{highly negative,
negative, neutral, positive, highly positive}~\cite{SemEval:2016:task4,SemEval:2017:task4}.
For the six NTCIR data sets,
the classes are five-point scale dialogue quality ratings ($-2, -1, 0, 1, 2$)
based on three different viewpoints, namely,
\emph{A-score} (task accomplishment),
\emph{E-score} (dialogue effectiveness), and
\emph{S-score} (customer satisfaction)~\cite{zeng19,zeng20}.
Hence, for example, DialEval-1 DQ-A
is the data set containing the gold and estimated probability distributions for the A-score estimation ``subsubtask'' of the NTCIR-15 DialEval-1 task.
The NTCIR dialogue data were provided in both Chinese and English (manually translated from the original Chinese text) to the participants, and the participants were allowed to submit Chinese and/or English runs.
On the other hand, the gold distributions were constructed solely based on the original Chinese dialogues. Hence,
both Chinese and English runs are evaluated using the same gold distributions.
The gold distributions of the STC-3 and DialEval-1 data were constructed based on votes from 19 and 20 assessors for each dialogue, respectively~\cite{zeng19}.

It should be noted that the NTCIR data sets are larger than the SemEval data sets, especially in terms of test data sample size.
Hence, our results with the NTCIR data sets may be more reliable than those with the SemEval data sets.


\section{System Ranking Agreement of Different Quantification Measures}

In this section, we compare the system rankings according to different measures, using Kendall's $\tau$~\cite{kendall62} with 95\% confidence intervals~\cite{long97}.
Tables~\ref{t:seme16-agreement}-\ref{t:deval1s-average}
show the results with each of the eight ordinal quantification data sets.
``Average similarity with other measures''
is computed as the average of all $\tau$'s involving a particular measure.
As a summary of these tables,
Table~\ref{t:average-agreement} and~\ref{t:average-average}
shows the average $\tau$'s across the eight data sets
and the average of the ``Average similarity with other measures,'' respectively.

The trends are very similar across the data sets and 
Table~\ref{t:average-average} serves as the summary of the results.
The following can be observed from this table:
\begin{itemize}
\item DNKT is the clear outlier measure: its system rankings are substantially different 
from the rankings according to other measures.
This is not surprising, as DNKT only cares about the PPP and ignores 
the absolute value of each probability.
\item While it does not quite stand out as much as DNKT,
NMD is also an outlier. RNOD and its variants 
behave relatively similarly to the nominal quantification measures (i.e., JSD, NVD, and RNSS).
This generalises the results of \citet{sakai21acl},
which showed that the property of RNOD lies
somewhere between NMD and the nominal quantification measures.
\end{itemize}

Tables~\ref{t:combining-semeval}-\ref{t:combining-dialeval1}
show how the combined measures DNKT\_JSD, DNKT\_NMD, and DNKT\_RNOD
behave in terms of system ranking when compared to their component measures.
As a summary,
Table~\ref{t:combining-average} shows the average $\tau$'s over the eight data sets.
It can be observed that each combined measure inherits the properties of both component measures,
which is not surprising.

\begin{table}[t]
\centering
\caption{System ranking agreement in terms of Kendall's $\tau$ with 95\%CIs (Sem16T4E, 12 runs).
}\label{t:seme16-agreement}\vspace*{-2mm}
\begin{scriptsize}
\begin{tabular}{c|ccccccccc}
\toprule
		&RNADW		&RNOD		&RNADW2	&RNOD2		&NVD		&RNSS		&JSD					&DNKT\\
\hline		
NMD	&0.848 		&0.909 		&0.818 		&0.848 		&0.939 		&0.818 		&0.909 					&0.667\\ 
		&[0.659, 0.936]&[0.787, 0.963]&[0.600, 0.923]&[0.659, 0.936]&[0.854, 0.975]	&[0.600, 0.923]	&[0.787, 0.963]	&[0.334, 0.852]\\
\hline
RN-	& - 			&0.939 		&0.970 		&1 			&0.848 		&0.970 		&0.818 					&0.576\\ 
ADW		&			&[0.854, 0.975]&[0.927, 0.988]&[1,1]		&[0.659, 0.936]&[0.927, 0.988]&[0.600, 0.923]			&[0.196, 0.806]\\
\hline
RNOD	& - 			& - 			&0.909 		&0.939 		&0.909 		&0.909 		&0.879 					&0.576\\ 
		&			&			&[0.787, 0.963]&[0.854, 0.975]&[0.787, 0.963]&[0.787, 0.963]&[0.723, 0.950]			&[0.196, 0.806]\\
\hline
RN-	& -			& - 			& - 			&0.970 		&0.818 		&1	 		&0.788 					&0.545\\ 
ADW2		&			&			&			&[0.927, 0.988]&[0.600, 0.923]&[1,1]		&[0.543, 0.909]			&[0.152, 0.789]\\
\hline
RN-	& -			& - 			& -			& -			&0.848 		&0.970 		&0.818 					&0.576\\ 
OD2		&			&			&			&			&[0.659, 0.936]&[0.927, 0.988]&[0.600, 0.923]			&[0.196, 0.806]\\
\hline
NVD		& -			& -			& - 			& - 			& -			&0.818 		&0.970 					&0.667\\ 
		&			&			&			&			&			&[0.600, 0.923]&[0.927, 0.988]			&[0.334, 0.852]\\
\hline
RNSS	& - 			& -  		& -			& -			& -			& - 			&0.788 					&0.545\\ 
		&			&			&			&			&			&			&[0.543, 0.909]			&[0.152, 0.789]\\
\hline
JSD		& -			& -			& -			& -			& -			& - 			& - 						&0.636\\
		& 			&			&			&			&			&			&						&[0.285, 0.837]\\
\bottomrule
\end{tabular}
\end{scriptsize}

\centering
\caption{Average similarity with other measures based on Table~\ref{t:seme16-agreement} (Sem16T4E).
}\label{t:seme16-average}\vspace*{-2mm}
\begin{scriptsize}
\begin{tabular}{c|r}
\toprule
Measure		&Average $\tau$\\
\hline
NMD		&0.845\\
RNADW		&0.871\\ 
RNOD		&0.871\\ 
RNADW2		&0.852\\ 
RNOD2		&0.871\\ 
NVD			&0.852\\ 
RNSS		&0.852\\ 
JSD			&0.826\\ 
DNKT		&0.599\\ 
\bottomrule
\end{tabular}
\end{scriptsize}
\end{table}

\begin{table}[t]
\centering
\caption{System ranking agreement in terms of Kendall's $\tau$  with 95\%CIs (Sem17T4E, 14 runs).
}\label{t:seme17-agreement}\vspace*{-2mm}
\begin{scriptsize}
\begin{tabular}{c|ccccccccc}
\toprule
		&RNADW		&RNOD		&RNADW2	&RNOD2		&NVD		&RNSS		&JSD			&DNKT\\
\hline		
NMD	&0.802 		&0.868 		&0.802 		&0.802 		&0.890 		&0.780 		&0.912 			&0.670\\ 
		&[0.601, 0.908]&[0.724, 0.940]&[0.601, 0.908]&[0.601, 0.908]&[0.767, 0.950]&[0.562, 0.897]&[0.811, 0.960]	&[0.381, 0.840]\\
\hline
RN-	& -			&0.934 		&0.956 		&0.956 		&0.912 		&0.978 		&0.890 			&0.692\\ 
ADW		&			&[0.856, 0.970]&[0.903, 0.980]&[0.903, 0.980]&[0.811, 0.960]&[0.951, 0.990]&[0.767, 0.950]	&[0.415, 0.851]\\
\hline
RNOD	& -	 		& -			&0.934 		&0.934 		&0.978 		&0.912 		&0.956 			&0.714\\ 
		&			&			&[0.856, 0.970]&[0.856, 0.970]&[0.951, 0.990]&[0.811, 0.960]&[0.903, 0.980]	&[0.451, 0.863]\\
\hline
RN-	& -			& -			& - 			&1 		&0.912 		&0.978 		&0.890 				&0.736\\ 
ADW2		&			&			&			&[1,1]	&[0.811, 0.960]&[0.951, 0.990]&[0.767, 0.950]		&[0.487, 0.874]\\
\hline
RN-	& -			& -			& -			& -			&0.912 		&0.978 		&0.890 			&0.736\\ 
OD2		&			&			&			&			&[0.811, 0.960]&[0.951, 0.990]&[0.767, 0.950]	&[0.487, 0.874]\\
\hline
NVD		& -			& - 			& -			& -			& -			&0.890 		&0.978 			&0.692\\
		&			&			&			&			&			&[0.767, 0.950]&[0.951, 0.990]	&[0.415, 0.851]\\
\hline
RNSS	& -			& -			& -			& -			& -			& -			&0.868 			&0.714\\ 
		&			&			&			&			&			&			&[0.724, 0.940]	&[0.451, 0.863]\\
\hline
JSD		& -			& -			& -			& -			& -			& -			& -				&0.670\\ 
		&			&			&			&			&			&			&				&[0.381, 0.840]\\
\bottomrule
\end{tabular}
\end{scriptsize}

\centering
\caption{Average similarity with other measures based on Table~\ref{t:seme17-agreement} (Sem17T4E).
}\label{t:seme17-average}\vspace*{-2mm}
\begin{scriptsize}
\begin{tabular}{c|r}
\toprule
Measure		&Average $\tau$\\
\hline
NMD		&0.816\\
RNADW		&0.890\\ 
RNOD		&0.904\\ 
RNADW2		&0.901\\ 
RNOD2		&0.901\\ 
NVD			&0.896\\ 
RNSS		&0.887\\ 
JSD			&0.882\\ 
DNKT		&0.703\\ 
\bottomrule
\end{tabular}
\end{scriptsize}
\end{table}

\begin{table}[t]
\centering
\caption{System ranking agreement in terms of Kendall's $\tau$  with 95\%CIs (STC-3 DQ-A, 19 runs).
}\label{t:stc3a-agreement}\vspace*{-2mm}
\begin{scriptsize}
\begin{tabular}{c|ccccccccc}
\toprule
		&RNADW		&RNOD		&RNADW2	&RNOD2		&NVD		&RNSS		&JSD			&DNKT\\
\hline		
NMD	&0.801 		&0.830 		&0.795 		&0.830 		&0.819 		&0.784 		&0.901 			&0.620\\
		&[0.645, 0.893]&[0.693, 0.909]&[0.635, 0.889]&[0.693, 0.909]&[0.675, 0.903]&[0.618, 0.883]&[0.815, 0.948]	&[0.372, 0.785]\\
\hline
RN-	& -			&0.953 		&0.977 		&0.953 		&0.953 		&0.965 		&0.883 			&0.556\\ 
ADW		&			&[0.910, 0.976]&[0.956, 0.988]&[0.910, 0.976]&[0.910, 0.976]&[0.933, 0.982]&[0.784, 0.938]	&[0.284, 0.745]\\
\hline
RNOD	& -			& -			&0.947 		&0.982 		&0.947 		&0.936 		&0.889 			&0.561\\ 
		&			&			&[0.899, 0.972]&[0.965, 0.991]&[0.899, 0.972]&[0.879, 0.967]&[0.794, 0.942]	&[0.291, 0.748]\\
\hline
RN-	& -			& -			& -			&0.947 		&0.959 		&0.971 		&0.877 			&0.550\\ 
ADW2		&			&			&			&[0.899, 0.972]&[0.921, 0.979]&[0.944, 0.985]&[0.773, 0.935]	&[0.276, 0.741]\\
\hline
RN-	& -			& -			& -			& -			&0.947 		&0.936 		&0.889 			&0.561\\ 
OD2		&			&			&			&			&[0.899, 0.972]&[0.879, 0.967]&[0.794, 0.942]	&[0.291, 0.748]\\
\hline
NVD		& -			& -			& -			& -			& -			&0.947 		&0.901 			&0.550\\ 
		&			&			&			&			&			&[0.899, 0.972]&[0.815, 0.948]	&[0.276, 0.741]\\
\hline
RNSS	& -			& -			& -			& - 			& - 			& -			&0.865 			&0.538\\ 
		&			&			&			&			&			&			&[0.752, 0.929]	&[0.261, 0.733]\\
\hline
JSD		& -			& -			& -			& -			& - 			& -			& -				&0.585\\ 
		&			&			&			&			&			&			&				&[0.323, 0.764]\\
\bottomrule
\end{tabular}
\end{scriptsize}

\centering
\caption{Average similarity with other measures based on Table~\ref{t:stc3a-agreement} (STC-3 DQ-A).
}\label{t:stc3a-average}\vspace*{-2mm}
\begin{scriptsize}
\begin{tabular}{c|r}
\toprule
Measure		&Average $\tau$\\
\hline 
NMD		&0.798\\ 
RNADW		&0.880\\  
RNOD		&0.881\\  
RNADW2		&0.878\\  
RNOD2		&0.881\\  
NVD			&0.878\\  
RNSS		&0.868\\  
JSD			&0.849\\  
DNKT		&0.565\\  
\bottomrule
\end{tabular}
\end{scriptsize}
\end{table}

\begin{table}[t]
\centering
\caption{System ranking agreement in terms of Kendall's $\tau$  with 95\%CIs (STC-3 DQ-E, 19 runs).
}\label{t:stc3e-agreement}\vspace*{-2mm}
\begin{scriptsize}
\begin{tabular}{c|ccccccccc}
\toprule
		&RNADW		&RNOD		&RNADW2	&RNOD2		&NVD		&RNSS		&JSD		&DNKT\\
\hline	
NMD	&0.836 		&0.871 		&0.836 		&0.871 		&0.848 		&0.836 		&0.906 		&0.322\\ 
		&[0.703, 0.913]&[0.763, 0.932]&[0.703, 0.913]&[0.763, 0.932]&[0.723, 0.919]&[0.703, 0.913]&[0.824, 0.951]&[$-$0.001, 0.584]\\
\hline
RN-	& - 			&0.947 		&0.982 		&0.947 		&0.971 		&0.982 		&0.889 		&0.433\\ 
ADW		&			&[0.899, 0.972]&[0.965, 0.991]&[0.899, 0.972]&[0.944, 0.985]&[0.965, 0.991]&[0.794, 0.942]&[0.128, 0.663]\\
\hline
RNOD	& - 			& -			&0.947 		&0.982 		&0.959 		&0.947 		&0.924 		&0.421\\ 
		&			&			&[0.899, 0.972]&[0.965, 0.991]&[0.921, 0.979]&[0.899, 0.972]&[0.857, 0.960]&[0.114, 0.655]\\
\hline
RN-	& - 			& - 			& -			&0.947 		&0.971 		&0.982 		&0.889 		&0.433\\ 
ADW2		&			&			&			&[0.899, 0.972]&[0.944, 0.985]&[0.965, 0.991]&[0.794, 0.942]&[0.128, 0.663]\\
\hline
RN-	& -			& -			& -			& -			&0.959 		&0.947 		&0.924 		&0.421\\ 
OD2		&			&			&			&			&[0.921, 0.979]&[0.899, 0.972]&[0.857, 0.960]&[0.114, 0.655]\\
\hline
NVD		& - 			& -			& -			& -			& -			&0.971 		&0.901 		&0.421\\ 
		&			&			&			&			&			&[0.944, 0.985]&[0.815, 0.948]&[0.114, 0.655]\\
\hline
RNSS	& - 			& -			& -			& -			& -			& -			&0.889 		&0.433\\ 
		&			&			&			&			&			&			&[0.794, 0.942]&[0.128, 0.663]\\
\hline
JSD		& -			& -			& -			& -			& -			& -			& -			&0.363\\ 
		&			&			&			&			&			&			&			&[0.046, 0.614]\\
\bottomrule
\end{tabular}
\end{scriptsize}

\centering
\caption{Average similarity with other measures based on Table~\ref{t:stc3e-agreement} (STC-3 DQ-E).
}\label{t:stc3e-average}\vspace*{-2mm}
\begin{scriptsize}
\begin{tabular}{c|r}
\toprule
Measure		&Average $\tau$\\
\hline
NMD		&0.791\\
RNADW		&0.873\\
RNOD		&0.875\\ 
RNADW2		&0.873\\ 
RNOD2		&0.875\\ 
NVD			&0.875\\ 
RNSS		&0.873\\ 
JSD			&0.836\\ 
DNKT		&0.406\\ 
\bottomrule
\end{tabular}
\end{scriptsize}
\end{table}

\begin{table}[t]
\centering
\caption{System ranking agreement in terms of Kendall's $\tau$  with 95\%CIs (STC-3 DQ-S, 19 runs).
}\label{t:stc3s-agreement}\vspace*{-2mm}
\begin{scriptsize}
\begin{tabular}{c|ccccccccc}
\toprule
		&RNADW		&RNOD		&RNADW2	&RNOD2		&NVD		&RNSS		&JSD		&DNKT\\
\hline		
NMD	&0.801 		&0.819 		&0.778 		&0.801 		&0.819 		&0.795 		&0.877 		&0.667\\ 
		&[0.645, 0.893]&[0.675, 0.903]&[0.608, 0.880]&[0.645, 0.893]&[0.675, 0.903]&[0.635, 0.889]&[0.773, 0.935]&[0.439, 0.814]\\
\hline
RN-	& -			&0.971 		&0.965 		&0.953 		&0.947 		&0.971 		&0.865 		&0.620\\ 
ADW		&			&[0.944, 0.985]&[0.933, 0.982]&[0.910, 0.976]&[0.899, 0.972]&[0.944, 0.985]&[0.752, 0.929]&[0.372, 0.785]\\
\hline
RNOD	& -			& -			&0.947 		&0.959		&0.965 		&0.965 		&0.883 		&0.626\\ 
		&			&			&[0.899, 0.972]&[0.921, 0.979]&[0.933, 0.982]&[0.933, 0.982]&[0.784, 0.938]&[0.380, 0.789]\\
\hline
RN-	& -			& -			& -			&0.953 		&0.924 		&0.947 		&0.842 		&0.608\\ 
ADW2		&			&			&			&[0.910, 0.976]&[0.857, 0.960]&[0.899, 0.972]&[0.713, 0.916]&[0.355, 0.778]\\
\hline
RN-	& -			& -			& -			& -			&0.947 		&0.959 		&0.865 		&0.620\\ 
OD2		&			&			&			&			&[0.899, 0.972]&[0.921, 0.979]&[0.752, 0.929]&[0.372, 0.785]\\
\hline
NVD		& -			& -			& -			& -			& -			&0.942 		&0.883 		&0.614\\ 
		&			&			&			&			&			&[0.890, 0.970]&[0.784, 0.938]&[0.363, 0.782]\\
\hline
RNSS	& - 			& -			& -			& -			& -			& -			&0.860 		&0.614\\ 
		&			&			&			&			&			&			&[0.744, 0.926]&[0.363, 0.782]\\
\hline
JSD		& -			& -			& -			& -			& -			& -			& -			&0.649\\ 
		&			&			&			&			&			&			&			&[0.413, 0.803]\\
\bottomrule
\end{tabular}
\end{scriptsize}

\centering
\caption{Average similarity with other measures based on Table~\ref{t:stc3s-agreement} (STC-3 DQ-S).
}\label{t:stc3s-average}\vspace*{-2mm}
\begin{scriptsize}
\begin{tabular}{c|r}
\toprule
Measure		&Average $\tau$\\
\hline
NMD		&0.795\\
RNADW		&0.887\\
RNOD		&0.892\\ 
RNADW2		&0.871\\ 
RNOD2		&0.882\\ 
NVD			&0.880\\ 
RNSS		&0.882\\ 
JSD			&0.841\\ 
DNKT		&0.627\\ 
\bottomrule
\end{tabular}
\end{scriptsize}
\end{table}

\begin{table}[t]
\centering
\caption{System ranking agreement in terms of Kendall's $\tau$  with 95\%CIs (DialEval-1 DQ-A, 22 runs).
}\label{t:deval1a-agreement}\vspace*{-2mm}
\begin{scriptsize}
\begin{tabular}{c|ccccccccc}
\toprule
		&RNADW		&RNOD		&RNADW2	&RNOD2		&NVD		&RNSS		&JSD		&DNKT\\
\hline	
NMD	&0.680 		&0.710 		&0.680 		&0.667 		&0.671 		&0.632 		&0.706 		&0.671\\ 
		&[0.481, 0.813]&[0.524, 0.831]&[0.481, 0.813]&[0.462, 0.804]&[0.468, 0.807]&[0.413, 0.782]&[0.518, 0.829]&[0.468, 0.807]\\
\hline
RN-	& -			&0.944 		&0.974 		&0.978 		&0.983 		&0.935 		&0.931 		&0.429\\ 
ADW		&			&[0.899, 0.969]&[0.953, 0.986]&[0.960, 0.988]&[0.969, 0.991]&[0.883, 0.964]&[0.876, 0.962]&[0.152, 0.643]\\
\hline
RNOD	& -			& -			&0.926 		&0.948 		&0.935 		&0.913 		&0.926 		&0.459\\ 
		&			&			&[0.868, 0.959]&[0.906, 0.971]&[0.883, 0.964]&[0.845, 0.952]&[0.868, 0.959]&[0.188, 0.665]\\
\hline
RN-	& -			& -			& -			&0.970 		&0.965 		&0.918 		&0.913 		&0.411\\ 
ADW2		&			&			&			&[0.945, 0.984]&[0.936, 0.981]&[0.854, 0.955]&[0.845, 0.952]&[0.131, 0.630]\\
\hline
RN-	& -			& -			& -			& -			&0.970 		&0.939 		&0.935 		&0.416\\ 
OD2		&			&			&			&			&[0.945, 0.984]&[0.890, 0.966]&[0.883, 0.964]&[0.137, 0.634]\\
\hline
NVD		& -			& -			& -			& -			& -			&0.926 		&0.922 		&0.420\\ 
		&			&			&			&			&			&[0.868, 0.959]&[0.861, 0.957]&[0.141, 0.637]\\
\hline
RNSS	& -			& -			& -			& -			& -			& -			&0.892 		&0.381\\ 
		&			&			&			&			&			&			&[0.810, 0.940]&[0.096, 0.609]\\
\hline
JSD		& -			& -			& -			& -			& -			& -			& -			&0.455\\ 
		&			&			&			&			&			&			&			&[0.183, 0.662]\\
\bottomrule
\end{tabular}
\end{scriptsize}

\centering
\caption{Average similarity with other measures based on Table~\ref{t:deval1a-agreement} (DialEval-1 DQ-A).
}\label{t:deval1a-average}\vspace*{-2mm}
\begin{scriptsize}
\begin{tabular}{c|r}
\toprule
Measure		&Average $\tau$\\
\hline
NMD		&0.677\\
RNADW		&0.857\\ 
RNOD		&0.845\\ 
RNADW2		&0.845\\ 
RNOD2		&0.853\\ 
NVD			&0.849\\
RNSS		&0.817\\ 
JSD			&0.835\\ 
DNKT		&0.455\\
\bottomrule
\end{tabular}
\end{scriptsize}
\end{table}

\begin{table}[t]
\centering
\caption{System ranking agreement in terms of Kendall's $\tau$  with 95\%CIs (DialEval-1 DQ-E, 22 runs).
}\label{t:deval1e-agreement}\vspace*{-2mm}
\begin{scriptsize}
\begin{tabular}{c|ccccccccc}
\toprule
		&RNADW		&RNOD		&RNADW2	&RNOD2		&NVD		&RNSS		&JSD		&DNKT\\
\hline	
NMD	&0.727 		&0.753 		&0.714 		&0.736 		&0.762 		&0.680 		&0.758 		&0.602\\ 
		&[0.549, 0.842]&[0.588, 0.858]&[0.530, 0.834]&[0.562, 0.847]&[0.602, 0.863]&[0.481, 0.813]&[0.595, 0.861]&[0.372, 0.762]\\
\hline
RN-	& -			&0.931 		&0.961 		&0.952 		&0.913 		&0.926 		&0.909 		&0.519\\ 
ADW		&			&[0.876, 0.962]&[0.929, 0.979]&[0.913, 0.974]&[0.845, 0.952]&[0.868, 0.959]&[0.839, 0.950]&[0.263. 0.707]\\
\hline
RNOD	& - 			& -			&0.900 		&0.957 		&0.948 		&0.909 		&0.970 		&0.571\\ 
		&			&			&[0.823, 0.944]&[0.922, 0.976]&[0.906, 0.971]&[0.839, 0.950]&[0.945, 0.984]&[0.331, 0.742]\\
\hline
RN-	& -			& -			& -			&0.926 		&0.883 		&0.939 		&0.879 		&0.489\\ 
ADW2		&			&			&			&[0.868, 0.959]&[0.795, 0.935]&[0.890, 0.966]&[0.788, 0.932]&[0.225, 0.686]\\
\hline
RN-	& -			& -			& -			& -			&0.931 		&0.935 		&0.944 		&0.537\\ 
OD2		&			&			&			&			&[0.876, 0.962]&[0.883, 0.964]&[0.899, 0.969]&[0.286, 0.719]\\
\hline
NVD		& -			& -			& -			& -			& -			&0.900 		&0.935 		&0.580\\ 
		&			&			&			&			&			&[0.823, 0.944]&[0.883, 0.964]&[0.343, 0.748]\\
\hline
RNSS	& -			& -			& -			& -			& -			& -			&0.905 		&0.489\\ 
		&			&			&			&			&			&			&[0.832, 0.947]&[0.225, 0.686]\\
\hline
JSD		& -			& -			& -			& -			& -			& -			& -			&0.576\\ 
		&			&			&			&			&			&			&			&[0.337, 0.745]\\
\bottomrule
\end{tabular}
\end{scriptsize}

\centering
\caption{Average similarity with other measures based on Table~\ref{t:deval1e-agreement} (DialEval-1 DQ-E).
}\label{t:deval1e-average}\vspace*{-2mm}
\begin{scriptsize}
\begin{tabular}{c|r}
\toprule
Measure		&Average $\tau$\\
\hline
NMD		&0.717\\
RNADW		&0.855\\
RNOD		&0.867\\ 
RNADW2		&0.836\\ 
RNOD2		&0.865\\ 
NVD			&0.857\\ 
RNSS		&0.835\\ 
JSD			&0.860\\ 
DNKT		&0.545\\ 
\bottomrule
\end{tabular}
\end{scriptsize}
\end{table}

\begin{table}[t]
\centering
\caption{System ranking agreement in terms of Kendall's $\tau$  with 95\%CIs (DialEval-1 DQ-S, 22 runs).
}\label{t:deval1s-agreement}\vspace*{-2mm}
\begin{scriptsize}
\begin{tabular}{c|ccccccccc}
\toprule
		&RNADW		&RNOD		&RNADW2	&RNOD2		&NVD		&RNSS		&JSD		&DNKT\\
\hline	
NMD	&0.584 		&0.576 		&0.584 		&0.571 		&0.576 		&0.550 		&0.610 		&0.680\\ 
		&[0.348, 0.750]&[0.337, 0.745]&[0.348, 0.750]&[0.331, 0.742]&[0.337, 0.745]&[0.303, 0.728]&[0.383, 0.768]&[0.481, 0.813]\\
\hline
RN-	& -			&0.983 		&0.991 		&0.978 		&0.983 		&0.957 		&0.948 		&0.532\\ 
ADW		&			&[0.969, 0.991]&[0.983, 0.995]&[0.960, 0.988]&[0.969, 0.991]&[0.922, 0.976]&[0.906, 0.971]&[0.280, 0.715]\\
\hline
RNOD	& -			& -			&0.983 		&0.987 		&0.974 		&0.965 		&0.939 		&0.524\\ 
		&			&			&[0.969, 0.991]&[0.976, 0.993]&[0.953, 0.986]&[0.936, 0.981]&[0.890, 0.966]&[0.270, 0.710]\\
\hline
RN-	& -			& -			& -			&0.978 		&0.983 		&0.957 		&0.948 		&0.532\\ 
ADW2		&			&			&			&[0.960, 0.988]&[0.969, 0.991]&[0.922, 0.976]&[0.906, 0.971]&[0.280, 0.715]\\
\hline
RN-	& -			& -			& -			& -			&0.970 		&0.970 		&0.935 		&0.519\\ 
OD2		&			&			&			&			&[0.945, 0.984]&[0.945, 0.984]&[0.883, 0.964]&[0.263, 0.707]\\
\hline
NVD		& -			& -			& -			& -			& -			&0.948 		&0.957 		&0.524\\ 
		&			&			&			&			&			&[0.906, 0.971]&[0.922, 0.976]&[0.270, 0.710]\\
\hline
RNSS	& -			& -			& -			& -			& -			& -			&0.913 		&0.498\\ 
		&			&			&			&			&			&			&[0.845, 0.952]&[0.237, 0.692]\\
\hline
JSD		& -			& -			& -			& -			& -			& -			& -			&0.541\\ 
		&			&			&			&			&			&			&			&[0.291, 0.722]\\
\bottomrule
\end{tabular}
\end{scriptsize}

\centering
\caption{Average similarity with other measures based on Table~\ref{t:deval1s-agreement} (DialEval-1 DQ-S).
}\label{t:deval1s-average}\vspace*{-2mm}
\begin{scriptsize}
\begin{tabular}{c|r}
\toprule
Measure		&Average $\tau$\\
\hline
NMD		&0.591\\
RNADW		&0.870\\
RNOD		&0.866\\ 
RNADW2		&0.870\\ 
RNOD2		&0.864\\ 
NVD			&0.864\\ 
RNSS		&0.845\\ 
JSD			&0.849\\ 
DNKT		&0.544\\ 
\bottomrule
\end{tabular}
\end{scriptsize}
\end{table}

\begin{table}[t]
\centering
\caption{System ranking agreement averaged over the eight data sets 
(Tables~\ref{t:seme16-agreement}, \ref{t:seme17-agreement}, \ref{t:stc3a-agreement}, \ref{t:stc3e-agreement}, \ref{t:stc3s-agreement}, \ref{t:deval1a-agreement}, \ref{t:deval1e-agreement}, \ref{t:deval1s-agreement}).
}\label{t:average-agreement}\vspace*{-2mm}
\begin{small}
\begin{tabular}{c|ccccccccc}
\toprule
		&RNADW		&RNOD		&RNADW2	&RNOD2		&NVD		&RNSS		&JSD		&DNKT\\
\hline	
NMD	&0.760 		&0.792 		&0.751 		&0.766 		&0.791 		&0.734 		&0.822 		&0.612\\ 
RNADW	& -			&0.950 		&0.972 		&0.965 		&0.939 		&0.961 		&0.892 		&0.545\\ 
RNOD	& -			& -			&0.937 		&0.961 		&0.952 		&0.932 		&0.921 		&0.557\\ 
RNADW2	& -			& -			& - 			&0.961 		&0.927 		&0.962 		&0.878 		&0.538\\ 
RNOD2	& -			& -			& -			& -			&0.936 		&0.954 		&0.900 		&0.548\\ 
NVD		& -			& -			& -			& -			& -			&0.918 		&0.931 		&0.559\\ 
RNSS	& -			& -			& -			& -			& -			& -			&0.873 		&0.527\\ 
JSD		& -			& -			& -			& -			& -			& -			& -			&0.559\\ 
\bottomrule
\end{tabular}
\end{small}

\centering
\caption{Average similarity with other measures further averaged over the eight data sets 
(Tables~\ref{t:seme16-average}, \ref{t:seme17-average}, \ref{t:stc3a-average}, \ref{t:stc3e-average}, \ref{t:stc3s-average}, \ref{t:deval1a-average}, \ref{t:deval1e-average}, \ref{t:deval1s-average}).
}\label{t:average-average}\vspace*{-2mm}
\begin{small}
\begin{tabular}{c|r}
\toprule
Measure		&Average $\tau$\\
\hline
NMD		&0.754\\
RNADW		&0.873\\
RNOD		&0.875\\ 
RNADW2		&0.866\\ 
RNOD2		&0.874\\ 
NVD			&0.869\\ 
RNSS		&0.857\\ 
JSD			&0.847\\ 
DNKT		&0.556\\ 
\bottomrule
\end{tabular}
\end{small}
\end{table}

\begin{table}[t]
\centering
\caption{The effect of combining DNKT and a quantification measure on system ranking  in terms of Kendall's $\tau$ with 95\%CIs (SemEval).
}\label{t:combining-semeval}
\begin{scriptsize}
\begin{tabular}{c|cc||c|cc||c|cc}
\toprule
\multicolumn{9}{c}{(a) Sem16T4E (12 runs)}\\
\hline
			&JSD	&DNKT\_JSD	&		&NMD	&DNKT\_NMD	&		&RNOD	&DNKT\_RNOD\\
\hline
DNKT		&0.636	&0.788		&DNKT	&0.667	&0.909		&DNKT	&0.576	&0.864\\
			&[0.285, 0.837]&[0.543, 0.909]& &[0.334, 0.852]&[0.787, 0.963]& &[0.196, 0.806]&[0.692, 0.943]\\
JSD			& -		&0.848		&NMD	& - 		&0.697		&RNOD	& -		&0.652\\
			& &[0.659, 0.936] 		&		& 		&[0.383, 0.867]&		&  		&[0.310, 0.845]\\
\hline
\multicolumn{9}{c}{(b) Sem17T4E (14 runs)}\\
\hline
			&JSD	&DNKT\_JSD	&		&NMD	&DNKT\_NMD	&		&RNOD	&DNKT\_RNOD\\
\hline
DNKT		&0.670	&0.714		&DNKT	&0.670	&0.758		&DNKT	&0.714	&0.824\\
			&[0.381, 0.840]&[0.451, 0.863]& &[0.381, 0.840]&[0.524, 0.886]& &[0.451, 0.863]&[0.641, 0.918]\\
JSD			& -		&0.956		&NMD	& -		&0.912		&RNOD	& -		&0.890\\
			&		&[0.903, 0.980]&		&		&[0.811, 0.960]&		&		&[0.767, 0.950]\\
\bottomrule
\end{tabular}
\end{scriptsize}
\end{table}

\begin{table}[t]
\centering
\caption{The effect of combining DNKT and a quantification measure on system ranking   in terms of Kendall's $\tau$ with 95\%CIs(STC-3).
}\label{t:combining-stc3}
\begin{scriptsize}
\begin{tabular}{c|cc||c|cc||c|cc}
\toprule
\multicolumn{9}{c}{(a) DQ-A (19 runs)}\\
\hline
			&JSD	&DNKT\_JSD	&		&NMD	&DNKT\_NMD	&		&RNOD	&DNKT\_RNOD\\
\hline
DNKT		&0.585	&0.678		&DNKT	&0.620	&0.678		&DNKT	&0.561	&0.690\\
		&[0.323, 0.764]&[0.455, 0.821]&	&[0.372, 0.785]&[0.455, 0.821]&	&[0.291, 0.748]&[0.473, 0.828]\\
JSD			& -		&0.842		&NMD	& -		&0.901		&RNOD	& -		&0.830\\
			&		&[0.713, 0.916]&		&		&[0.815, 0.948]&		&		&[0.693, 0.909]\\
\hline
\multicolumn{9}{c}{(b) DQ-E (19 runs)}\\
\hline
			&JSD	&DNKT\_JSD	&		&NMD	&DNKT\_NMD	&		&RNOD	&DNKT\_RNOD\\
\hline
DNKT		&0.363	&0.462		&DNKT	&0.322	&0.450		&DNKT	&0.421	&0.538\\
		&[0.046, 0.614]&[0.164, 0.683]&	&[$-$0.001, 0.584]&[0.149, 0.675]& &[0.114, 0.655]&[0.261, 0.733]\\
JSD			& -		&0.883		&NMD	& -		&0.842		&RNOD	& -		&0.842\\
			&		&[0.784, 0.938]&		&		&[0.713, 0.916]&		&		&[0.713, 0.916]\\
\hline
\multicolumn{9}{c}{(c) DQ-S (19 runs)}\\
\hline
			&JSD	&DNKT\_JSD	&		&NMD	&DNKT\_NMD	&		&RNOD	&DNKT\_RNOD\\
\hline
DNKT		&0.649	&0.772		&DNKT	&0.667	&0.760		&DNKT	&0.626	&0.807\\
		&[0.413, 0.803]&[0.598, 0.876]&	&[0.439, 0.814]&[0.579, 0.869]&	&[0.380, 0.789]&[0.655, 0.896]\\
JSD			& -		&0.865		&NMD	& -		&0.883		&RNOD	& -		&0.807\\
			&		&[0.752, 0.929]&		&		&[0.784, 0.938]&		&		&[0.655, 0.896]\\
\bottomrule
\end{tabular}
\end{scriptsize}
\end{table}

\begin{table}[t]
\centering
\caption{The effect of combining DNKT and a quantification measure on system ranking  in terms of Kendall's $\tau$ with 95\%CIs (DialEval-1).
}\label{t:combining-dialeval1}
\begin{scriptsize}
\begin{tabular}{c|cc||c|cc||c|cc}
\toprule
\multicolumn{9}{c}{(a) DQ-A (22 runs)}\\
\hline
			&JSD	&DNKT\_JSD	&		&NMD	&DNKT\_NMD	&		&RNOD	&DNKT\_RNOD\\
\hline
DNKT		&0.455	&0.610		&DNKT	&0.671	&0.948		&DNKT	&0.459	&0.874\\
		&[0.183, 0.662]&[0.383, 0.768]&	&[0.468, 0.807]&[0.906, 0.971]&	&[0.188, 0.665]&[0.780, 0.930]\\
JSD			& -		&0.835		&NMD	& -		&0.714		&RNOD	& -		&0.576\\
			&		&[0.716, 0.907]&		&		&[0.530, 0.834]&		&		&[0.337, 0.745]\\
\hline
\multicolumn{9}{c}{(b) DQ-E (22 runs)}\\
\hline
			&JSD	&DNKT\_JSD	&		&NMD	&DNKT\_NMD	&		&RNOD	&DNKT\_RNOD\\
\hline
DNKT		&0.576	&0.662		&DNKT	&0.602	&0.740		&DNKT	&0.571	&0.693\\
		&[0.337, 0.745]&[0.455, 0.801]&	&[0.372, 0.762]&[0.568, 0.850]&	&[0.331, 0.742]&[0.499, 0.821]\\
JSD			& -		&0.905		&NMD	& -		&0.835		&RNOD	& -		&0.853\\
			&		&[0.832, 0.947]&		&		&[0.716, 0.907]&		&		&[0.745, 0.917]\\
\hline
\multicolumn{9}{c}{(c) DQ-S (22 runs)}\\
\hline
			&JSD	&DNKT\_JSD	&		&NMD	&DNKT\_NMD	&		&RNOD	&DNKT\_RNOD\\
\hline
DNKT		&0.541	&0.636		&DNKT	&0.680	&0.861		&DNKT	&0.524	&0.758\\
		&[0.291, 0.722]&[0.419, 0.784]&	&[0.481, 0.813]&[0.758, 0.922]&	&[0.270, 0.710]&[0.595, 0.861]\\
JSD			& -		&0.896		&NMD	& -		&0.810		&RNOD	& -		&0.749\\
			&	&[0.816, 0.942]	&		&	&[0.676, 0.892]	&		&	&[0.582, 0.855]\\
\bottomrule
\end{tabular}
\end{scriptsize}
\end{table}

\begin{table}[t]
\centering
\caption{The effect of combining DNKT and a quantification measure on system ranking: average over the eight data sets (Tables~\ref{t:combining-semeval}-\ref{t:combining-dialeval1}).
}\label{t:combining-average}
\begin{small}
\begin{tabular}{c|cc||c|cc||c|cc}
\toprule
			&JSD	&DNKT\_JSD	&		&NMD	&DNKT\_NMD	&		&RNOD	&DNKT\_RNOD\\
\hline
DNKT		&0.559	&0.665		&DNKT	&0.612	&0.763		&DNKT	&0.557	&0.756\\
JSD			& -		&0.879		&NMD	& -		&0.824		&RNOD	& - 		&0.775\\
\bottomrule
\end{tabular}
\end{small}
\end{table}

\clearpage

\section{System Ranking Consistency of Each Quantification Measure}\label{ss:consistency-quantification}

Next, we discuss the system ranking consistency (i.e., robustness of the ranking to the choice of test data)
of the measures discussed above,
using the procedure described in \citet[Figures 1 and 2]{sakai21ecir}.
That is, for each data set, the test data is split in half
so that two system rankings are obtained according to Mean JSD etc.;
the similarity of the two rankings is quantified by Kendall's $\tau$;
The random splitting is carried out $B=1,000$ times so that Mean $\tau$'s are obtained.
Similarly, to consider situations where the test data sample sizes are small,
two disjoint test sets of size 10 are randomly sampled from the original test set (instead of splitting it in half),
and Mean $\tau$'s are obtained in the same way.
The randomised Tukey HSD test~\cite{sakai18book} at the significance level of $\alpha=0.05$
is used to discuss the statistical significance 
of the differences in system ranking consistency.

Tables~\ref{t:consistency-semeval}-\ref{t:consistency-deval1} show the system ranking consistency results.
For example, in Table~\ref{t:consistency-semeval}(a) where the 12 Sem16T4E systems are ranked,
JSD statistically significantly outperforms all other measures for both test set sizes (50 and 10). 
RNOD comes second,  statistically significantly outperforming nine other measures when the test set size is 50,
and seven other measures when the test set size is 10.
As a summary of the eight sets of results,
Table~\ref{t:consistency-average} shows the mean $\tau$'s averaged over these results.
In terms of system ranking consistency,
it can be observed from this summary table that:
\begin{itemize}
\item Among the nominal quantification measures,
NVD and JSD are the overall winners, although
RNSS is the top performer in
the ``full split'' experiments with STC-3 DQ-E and DQ-S (Table~\ref{t:consistency-stc3}).
\item Among the ordinal quantification measures
(i.e., NMD, RNOD, and the variants of RNOD),
RNOD is the winner.
In this sense, the variants of RNOD do not offer any benefit.
\item DNKT
performs worse than any other measure.
That is, the system ranking according to DNKT 
changes substantially depending on the choice of test data.
Because of this property, the combined measures
(DNKT\_JSD, DNKT\_NMD, DNKT\_RNOD) do not do well either.
\end{itemize}

\begin{table}[t]
\centering

\caption{Mean System Ranking Consistency $\tau$'s (SemEval data).
The symbols 
indicate ``statistically significantly outperforms
NMD ($\dagger$), 
RNADW ($\clubsuit$), 
RNOD ($\heartsuit$), 
RNADW2 ($\spadesuit$), 
RNOD2 ($\diamondsuit$), 
NVD ($\ast$), 
RNSS ($\star$), 
JSD ($\natural$), 
DNKT ($\circ$),
DNKT\_JSD ($\bigcirc$),
DNKT\_NMD ($\bigtriangleup$),
DNKT\_RNOD ($\bigtriangledown$)
at the 5\% significance level with a randomised Tukey HSD test,'' respectively.
}\label{t:consistency-semeval}
\begin{small}
\begin{tabular}{crr|crr}
\toprule
\multicolumn{6}{c}{(a) Sem16T4E}\\
\hline
\multicolumn{3}{c|}{Full split (50 vs. 50)} &\multicolumn{3}{c}{10 vs. 10}\\
\hline
JSD 			&0.9336&$\heartsuit\bigcirc\ast\spadesuit\diamondsuit\clubsuit\star\circ\bigtriangledown\dagger\bigtriangleup$&JSD 			&0.7711&$\heartsuit\bigcirc\ast\clubsuit\spadesuit\diamondsuit\star\dagger\bigtriangleup\bigtriangledown\circ$\\
RNOD 			&0.8470&$\ast\spadesuit\diamondsuit\clubsuit\star\circ\bigtriangledown\dagger\bigtriangleup$ &RNOD 			&0.7077&$\spadesuit\diamondsuit\star\dagger\bigtriangleup\bigtriangledown\circ$\\
DNKT\_JSD 		&0.8456&$\ast\spadesuit\diamondsuit\clubsuit\star\circ\bigtriangledown\dagger\bigtriangleup$ &DNKT\_JSD 		&0.7069&$\spadesuit\diamondsuit\star\dagger\bigtriangleup\bigtriangledown\circ$\\
NVD 			&0.8307&$\diamondsuit\clubsuit\star\circ\bigtriangledown\dagger\bigtriangleup$ &NVD 			&0.7049&$\spadesuit\diamondsuit\star\dagger\bigtriangleup\bigtriangledown\circ$\\
RNADW2 		&0.8229&$\circ\bigtriangledown\dagger\bigtriangleup$ &RNADW 			&0.7033&$\diamondsuit\star\dagger\bigtriangleup\bigtriangledown\circ$\\
RNOD2 			&0.8212&$\circ\bigtriangledown\dagger\bigtriangleup$ &RNADW2 		&0.6921&$\dagger\bigtriangleup\bigtriangledown\circ$\\
RNADW 			&0.8185&$\circ\bigtriangledown\dagger\bigtriangleup$ &RNOD2 			&0.6908&$\dagger\bigtriangleup\bigtriangledown\circ$\\
RNSS 			&0.8147&$\circ\bigtriangledown\dagger\bigtriangleup$ &RNSS 			&0.6898&$\dagger\bigtriangleup\bigtriangledown\circ$\\
DNKT 			&0.8040&$\bigtriangledown\dagger\bigtriangleup$ 		&NMD 			&0.6743\\
DNKT\_RNOD 	&0.7918& &DNKT\_NMD 		&0.6667\\
NMD 			&0.7881& &DNKT\_RNOD 	&0.6632\\
DNKT\_NMD 		&0.7833& &DNKT 			&0.6629\\
\hline
\multicolumn{6}{c}{(b) Sem17T4E}\\
\hline
\multicolumn{3}{c|}{Full split (63 vs. 62)} &\multicolumn{3}{c}{10 vs. 10}\\
\hline	
NMD 			&0.9046 &$\ast\natural\heartsuit\bigcirc\diamondsuit\clubsuit\spadesuit\bigtriangleup\star\bigtriangledown\circ$&NMD &0.7053&$\natural\bigcirc\ast\bigtriangleup\heartsuit\diamondsuit\clubsuit\spadesuit\star\bigtriangledown\circ$\\
NVD 			&0.8779 &$\natural\heartsuit\bigcirc\diamondsuit\clubsuit\spadesuit\bigtriangleup\star\bigtriangledown\circ$&JSD &0.6720&$\bigcirc\ast\bigtriangleup\heartsuit\diamondsuit\clubsuit\spadesuit\star\bigtriangledown\circ$\\
JSD 			&0.8668 &$\heartsuit\bigcirc\diamondsuit\clubsuit\spadesuit\bigtriangleup\star\bigtriangledown\circ$&DNKT\_JSD &0.6083&$\bigtriangleup\heartsuit\diamondsuit\clubsuit\spadesuit\star\bigtriangledown\circ$\\
RNOD 			&0.8256 &$\diamondsuit\clubsuit\spadesuit\bigtriangleup\star\bigtriangledown\circ$&NVD &0.6007&$\diamondsuit\clubsuit\spadesuit\star\bigtriangledown\circ$\\
DNKT\_JSD 		&0.8222 &$\diamondsuit\clubsuit\spadesuit\bigtriangleup\star\bigtriangledown\circ$&DNKT\_NMD &0.5895&$\diamondsuit\clubsuit\spadesuit\star\bigtriangledown\circ$\\
RNOD2 			&0.8107 &$\spadesuit\bigtriangleup\star\bigtriangledown\circ$&RNOD &0.5882&$\diamondsuit\clubsuit\spadesuit\star\bigtriangledown\circ$\\
RNADW 			&0.8015 &$\star\bigtriangledown\circ$&RNOD2 &0.5729&$\star\bigtriangledown\circ$\\
RNADW2 		&0.7956 &$\star\bigtriangledown\circ$&RNADW &0.5672&$\bigtriangledown\circ$\\
DNKT\_NMD 		&0.7943 &$\star\bigtriangledown\circ$&RNADW2 &0.5607&$\bigtriangledown\circ$\\
RNSS 			&0.7650 &$\circ$&RNSS &0.5569&$\bigtriangledown\circ$\\
DNKT\_RNOD 	&0.7558 &&DNKT\_RNOD &0.5110&$\circ$\\
DNKT 			&0.7504 &&DNKT &0.4530&\\
\bottomrule
\end{tabular}
\end{small}
\end{table}

\begin{table}[t]
\centering

\caption{Mean System Ranking Consistency $\tau$'s (STC-3 data).
The symbols 
indicate ``statistically significantly outperforms
NMD ($\dagger$), 
RNADW ($\clubsuit$), 
RNOD ($\heartsuit$), 
RNADW2 ($\spadesuit$), 
RNOD2 ($\diamondsuit$), 
NVD ($\ast$), 
RNSS ($\star$), 
JSD ($\natural$), 
DNKT ($\circ$),
DNKT\_JSD ($\bigcirc$),
DNKT\_NMD ($\bigtriangleup$),
DNKT\_RNOD ($\bigtriangledown$)
at the 5\% significance level with a randomised Tukey HSD test,'' respectively.
}\label{t:consistency-stc3}
\begin{small}
\begin{tabular}{crr|crr}
\toprule
\multicolumn{6}{c}{(a) STC-3 DQ-A}\\
\hline
\multicolumn{3}{c|}{Full split (195 vs. 195)} &\multicolumn{3}{c}{10 vs. 10}\\
\hline
RNADW2 	&0.7677 &$\clubsuit\heartsuit\bigtriangledown\bigcirc\natural\dagger\bigtriangleup\circ$&RNADW2 		&0.4479 &$\natural\bigcirc\bigtriangledown\dagger\bigtriangleup\circ$\\
RNSS 		&0.7662 &$\clubsuit\heartsuit\bigtriangledown\bigcirc\natural\dagger\bigtriangleup\circ$&RNSS 			&0.4472 &$\natural\bigcirc\bigtriangledown\dagger\bigtriangleup\circ$\\
RNOD2 		&0.7650 &$\clubsuit\heartsuit\bigtriangledown\bigcirc\natural\dagger\bigtriangleup\circ$&RNADW 		&0.4442 &$\natural\bigcirc\bigtriangledown\dagger\bigtriangleup\circ$\\
NVD 		&0.7647 &$\clubsuit\heartsuit\bigtriangledown\bigcirc\natural\dagger\bigtriangleup\circ$&RNOD2 		&0.4416 &$\natural\bigcirc\bigtriangledown\dagger\bigtriangleup\circ$\\
RNADW 		&0.7518 &$\bigtriangledown\bigcirc\natural\dagger\bigtriangleup\circ$&NVD 			&0.4393 &$\bigcirc\bigtriangledown\dagger\bigtriangleup\circ$\\
RNOD 		&0.7506 &$\bigtriangledown\bigcirc\natural\dagger\bigtriangleup\circ$&RNOD 			&0.4350 &$\bigtriangledown\dagger\bigtriangleup\circ$\\
DNKT\_RNOD &0.7373 &$\bigcirc\natural\dagger\bigtriangleup\circ$&JSD 			&0.4253 &$\dagger\bigtriangleup\circ$\\
DNKT\_JSD 	&0.7267 &$\natural\dagger\bigtriangleup\circ$&DNKT\_JSD 	&0.4248 &$\dagger\bigtriangleup\circ$\\
JSD 		&0.7071 &$\bigtriangleup\circ$&DNKT\_RNOD 	&0.4151 &$\dagger\bigtriangleup\circ$\\
NMD 		&0.7060 &$\bigtriangleup\circ$&NMD 			&0.3966 &$\circ$\\
DNKT\_NMD 	&0.6862 &$\circ$&DNKT\_NMD 	&0.3909 &$\circ$\\
DNKT 		&0.5616 &&DNKT 			&0.2402 &\\
\hline

\multicolumn{6}{c}{(b) STC-3 DQ-E}\\
\hline
\multicolumn{3}{c|}{Full split (195 vs. 195)} &\multicolumn{3}{c}{10 vs. 10}\\
\hline
RNSS 		&0.7853 &$\heartsuit\natural\dagger\bigtriangledown\bigcirc\bigtriangleup\circ$&NVD 			&0.4512 &$\bigtriangledown\bigcirc\bigtriangleup\dagger\circ$\\
NVD 		&0.7805 &$\natural\dagger\bigtriangledown\bigcirc\bigtriangleup\circ$&RNSS 			&0.4503 &$\bigtriangledown\bigcirc\bigtriangleup\dagger\circ$\\
RNADW 		&0.7790 &$\natural\dagger\bigtriangledown\bigcirc\bigtriangleup\circ$&RNADW 		&0.4478 &$\bigtriangledown\bigcirc\bigtriangleup\dagger\circ$\\
RNADW2 	&0.7777 &$\natural\dagger\bigtriangledown\bigcirc\bigtriangleup\circ$&RNADW2 		&0.4455 &$\bigtriangledown\bigcirc\bigtriangleup\dagger\circ$\\
RNOD2 		&0.7768 &$\natural\dagger\bigtriangledown\bigcirc\bigtriangleup\circ$&RNOD2 		&0.4454 &$\bigtriangledown\bigcirc\bigtriangleup\dagger\circ$\\
RNOD 		&0.7733 &$\natural\dagger\bigtriangledown\bigcirc\bigtriangleup\circ$&RNOD 			&0.4449 &$\bigtriangledown\bigcirc\bigtriangleup\dagger\circ$\\
JSD 		&0.7545 &$\bigtriangledown\bigcirc\bigtriangleup\circ$&JSD 			&0.4447 &$\bigtriangledown\bigcirc\bigtriangleup\dagger\circ$\\
NMD 		&0.7482 &$\bigtriangledown\bigcirc\bigtriangleup\circ$&DNKT\_RNOD 	&0.4286 &$\bigtriangleup\dagger\circ$\\
DNKT\_RNOD &0.7187 &$\bigcirc\bigtriangleup\circ$&DNKT\_JSD 	&0.4174 &$\circ$\\
DNKT\_JSD 	&0.6844 &$\bigtriangleup\circ$&DNKT\_NMD 	&0.4129 &$\circ$\\
DNKT\_NMD 	&0.6639 &$\circ$&NMD 			&0.4124 &$\circ$\\
DNKT 		&0.6205 &&DNKT 			&0.2671 &\\
\hline


\multicolumn{6}{c}{(c) STC-3 DQ-S}\\
\hline
\multicolumn{3}{c|}{Full split (195 vs. 195)} &\multicolumn{3}{c}{10 vs. 10}\\
\hline
RNSS 		&0.7556 &$\ast\diamondsuit\heartsuit\dagger\bigcirc\bigtriangledown\natural\bigtriangleup\circ$&RNADW2 		&0.4207 &$\dagger\heartsuit\bigtriangledown\bigtriangleup\circ$\\
RNADW2 	&0.7497 &$\diamondsuit\heartsuit\dagger\bigcirc\bigtriangledown\natural\bigtriangleup\circ$&RNSS 			&0.4193 &$\dagger\heartsuit\bigtriangledown\bigtriangleup\circ$\\
RNADW 		&0.7492 &$\diamondsuit\heartsuit\dagger\bigcirc\bigtriangledown\natural\bigtriangleup\circ$&RNADW 		&0.4179 &$\heartsuit\bigtriangledown\bigtriangleup\circ$\\
NVD 		&0.7413 &$\diamondsuit\heartsuit\dagger\bigcirc\bigtriangledown\natural\bigtriangleup\circ$&DNKT\_JSD 	&0.4150 &$\bigtriangleup\circ$\\
RNOD2 		&0.7273 &$\dagger\bigcirc\bigtriangledown\natural\bigtriangleup\circ$&RNOD2 		&0.4102 &$\bigtriangleup\circ$\\
RNOD 		&0.7166 &$\dagger\bigcirc\bigtriangledown\natural\bigtriangleup\circ$&JSD 			&0.4071 &$\bigtriangleup\circ$\\
NMD 		&0.6935 &$\bigtriangledown\natural\bigtriangleup\circ$&NVD 			&0.4070 &$\bigtriangleup\circ$\\
DNKT\_JSD 	&0.6852 &$\bigtriangleup\circ$&NMD 			&0.4046 &$\bigtriangleup\circ$\\
DNKT\_RNOD &0.6807 &$\bigtriangleup\circ$&RNOD 			&0.4033 &$\circ$\\
JSD 		&0.6743 &$\bigtriangleup\circ$&DNKT\_RNOD 	&0.4031 &$\circ$\\
DNKT\_NMD 	&0.6529 &$\circ$&DNKT\_NMD 	&0.3899 &$\circ$\\
DNKT 		&0.4944 &&DNKT 			&0.2359 &\\
\bottomrule

\end{tabular}
\end{small}
\end{table}

\begin{table}[t]
\centering
\caption{Mean System Ranking Consistency $\tau$'s (DialEval-1 data).
The symbols 
indicate ``statistically significantly outperforms
NMD ($\dagger$), 
RNADW ($\clubsuit$), 
RNOD ($\heartsuit$), 
RNADW2 ($\spadesuit$), 
RNOD2 ($\diamondsuit$), 
NVD ($\ast$), 
RNSS ($\star$), 
JSD ($\natural$), 
DNKT ($\circ$),
DNKT\_JSD ($\bigcirc$),
DNKT\_NMD ($\bigtriangleup$),
DNKT\_RNOD ($\bigtriangledown$)
at the 5\% significance level with a randomised Tukey HSD test,'' respectively.
}\label{t:consistency-deval1}
\begin{small}
\begin{tabular}{crr|crr}
\toprule
\multicolumn{6}{c}{(a) DialEval-1 DQ-A}\\
\hline
\multicolumn{3}{c|}{Full split (150 vs. 150)} &\multicolumn{3}{c}{10 vs. 10}\\
\hline
RNOD2 			&0.9019 &$\star\ast\natural\bigtriangleup\circ\bigtriangledown\bigcirc\dagger$&JSD 			&0.5495 &$\circ\heartsuit\ast\spadesuit\clubsuit\diamondsuit\bigtriangledown\bigcirc\bigtriangleup\star\dagger$\\
RNOD 			&0.8999 &$\star\ast\natural\bigtriangleup\circ\bigtriangledown\bigcirc\dagger$&DNKT 			&0.5295 &$\heartsuit\ast\spadesuit\clubsuit\diamondsuit\bigtriangledown\bigcirc\bigtriangleup\star\dagger$\\
RNADW 			&0.8983 &$\star\ast\natural\bigtriangleup\circ\bigtriangledown\bigcirc\dagger$&RNOD 			&0.4989 &$\spadesuit\clubsuit\diamondsuit\bigtriangledown\bigcirc\bigtriangleup\star\dagger$\\
RNADW2 		&0.8954 &$\star\ast\natural\bigtriangleup\circ\bigtriangledown\bigcirc\dagger$&NVD 			&0.4890 &$\bigtriangledown\bigcirc\bigtriangleup\star\dagger$\\
RNSS 			&0.8760 &$\bigtriangleup\circ\bigtriangledown\bigcirc\dagger$&RNADW2 		&0.4809 &$\bigtriangledown\bigcirc\bigtriangleup\star\dagger$\\
NVD 			&0.8738 &$\bigtriangleup\circ\bigtriangledown\bigcirc\dagger$&RNADW 		&0.4797 &$\bigtriangledown\bigcirc\bigtriangleup\star\dagger$\\
JSD 			&0.8699 &$\bigtriangleup\circ\bigtriangledown\bigcirc\dagger$&RNOD2 		&0.4753 &$\bigtriangledown\bigcirc\bigtriangleup\star\dagger$\\
DNKT\_NMD 		&0.8452 &$\bigtriangledown\bigcirc\dagger$&DNKT\_RNOD 	&0.4579 &$\dagger$\\
DNKT 			&0.8404 &$\bigtriangledown\bigcirc\dagger$&DNKT\_JSD 	&0.4568 &$\dagger$\\
DNKT\_RNOD 	&0.8184 &$\bigcirc\dagger$&DNKT\_NMD 	&0.4541 &$\dagger$\\
DNKT\_JSD 		&0.7683 &$\dagger$&RNSS 			&0.4474 &$\dagger$\\
NMD 			&0.7085 &&NMD 			&0.4149 &\\
\hline
\multicolumn{6}{c}{(b) DialEval-1 DQ-E}\\
\hline
\multicolumn{3}{c|}{Full split (150 vs. 150)} &\multicolumn{3}{c}{10 vs. 10}\\
\hline
NMD 			&0.8568 &$\natural\star\bigcirc\ast\bigtriangledown\heartsuit\diamondsuit\clubsuit\circ\spadesuit\bigtriangleup$&JSD 			&0.6150 &$\heartsuit\diamondsuit\clubsuit\star\ast\spadesuit\bigcirc\circ\dagger\bigtriangledown\bigtriangleup$\\
JSD 			&0.8329 &$\star\bigcirc\ast\bigtriangledown\heartsuit\diamondsuit\clubsuit\circ\spadesuit\bigtriangleup$&RNOD 			&0.6010 &$\diamondsuit\clubsuit\star\ast\spadesuit\bigcirc\circ\dagger\bigtriangledown\bigtriangleup$\\
RNSS 			&0.8265 &$\ast\bigtriangledown\heartsuit\diamondsuit\clubsuit\circ\spadesuit\bigtriangleup$&RNOD2 		&0.5869 &$\spadesuit\bigcirc\circ\dagger\bigtriangledown\bigtriangleup$\\
DNKT\_JSD 		&0.8248 &$\ast\bigtriangledown\heartsuit\diamondsuit\clubsuit\circ\spadesuit\bigtriangleup$&RNADW 		&0.5856 &$\bigcirc\circ\dagger\bigtriangledown\bigtriangleup$\\
NVD 			&0.8100 &$\heartsuit\diamondsuit\clubsuit\circ\spadesuit\bigtriangleup$&RNSS 			&0.5849 &$\bigcirc\circ\dagger\bigtriangledown\bigtriangleup$\\
DNKT\_RNOD 	&0.8071 &$\clubsuit\circ\spadesuit\bigtriangleup$&NVD 			&0.5834 &$\bigcirc\circ\dagger\bigtriangledown\bigtriangleup$\\
RNOD 			&0.8041 &$\circ\spadesuit\bigtriangleup$&RNADW2 		&0.5735 &$\bigcirc\circ\dagger\bigtriangledown\bigtriangleup$\\
RNOD2 			&0.8021 &$\circ\spadesuit\bigtriangleup$&DNKT\_JSD 	&0.5566 &$\circ\dagger\bigtriangledown\bigtriangleup$\\
RNADW 			&0.8013 &$\circ\spadesuit\bigtriangleup$&DNKT 			&0.5164 &$\dagger\bigtriangledown\bigtriangleup$\\
DNKT 			&0.7926 &$\bigtriangleup$&NMD 			&0.4933 &$\bigtriangledown\bigtriangleup$\\
RNADW2 		&0.7911 &$\bigtriangleup$&DNKT\_RNOD 	&0.4725 &$\bigtriangleup$\\
DNKT\_NMD 		&0.7797 &&DNKT\_NMD 	&0.4375 &\\
\hline


\multicolumn{6}{c}{(c) DialEval-1 DQ-S}\\
\hline
\multicolumn{3}{c|}{Full split (150 vs. 150)} &\multicolumn{3}{c}{10 vs. 10}\\
\hline
RNOD2 			&0.9007 &$\heartsuit\spadesuit\ast\circ\bigtriangledown\bigtriangleup\dagger$&JSD 			&0.5714 &$\heartsuit\circ\bigcirc\diamondsuit\ast\clubsuit\spadesuit\star\bigtriangledown\bigtriangleup\dagger$\\
RNSS 			&0.8972 &$\heartsuit\spadesuit\ast\circ\bigtriangledown\bigtriangleup\dagger$&RNOD 			&0.5382 &$\bigcirc\diamondsuit\ast\clubsuit\spadesuit\star\bigtriangledown\bigtriangleup\dagger$\\
DNKT\_JSD 		&0.8940 &$\ast\circ\bigtriangledown\bigtriangleup\dagger$&DNKT 			&0.5252 &$\spadesuit\star\bigtriangledown\bigtriangleup\dagger$\\
RNADW 			&0.8926 &$\ast\circ\bigtriangledown\bigtriangleup\dagger$&DNKT\_JSD 	&0.5231 &$\star\bigtriangledown\bigtriangleup\dagger$\\
JSD 			&0.8925 &$\ast\circ\bigtriangledown\bigtriangleup\dagger$&RNOD2		 	&0.5203&$\star\bigtriangledown\bigtriangleup\dagger$\\
RNOD 			&0.8882 &$\ast\circ\bigtriangledown\bigtriangleup\dagger$&NVD 			&0.5145 &$\bigtriangledown\bigtriangleup\dagger$\\
RNADW2 		&0.8879 &$\ast\circ\bigtriangledown\bigtriangleup\dagger$&RNADW 		&0.5127 &$\bigtriangledown\bigtriangleup\dagger$\\
NVD 			&0.8615 &$\bigtriangledown\bigtriangleup\dagger$&RNADW2 		&0.5108 &$\bigtriangledown\bigtriangleup\dagger$\\
DNKT 			&0.8550 &$\bigtriangledown\bigtriangleup\dagger$&RNSS 			&0.5054 &$\bigtriangledown\bigtriangleup\dagger$\\
DNKT\_RNOD 	&0.8170 &$\bigtriangleup\dagger$&DNKT\_RNOD 	&0.4780 &$\bigtriangleup\dagger$\\
DNKT\_NMD 		&0.7869 &$\dagger$&DNKT\_NMD 	&0.4480 &$\dagger$\\
NMD 			&0.7360 &&NMD 			&0.4120 &\\
\bottomrule
\end{tabular}
\end{small}
\end{table}

\begin{table}[t]
\centering

\caption{Mean System Ranking Consistency averaged over the eight data sets.
}\label{t:consistency-average}
\begin{small}
\begin{tabular}{cr|cr}
\toprule
\multicolumn{2}{c|}{Full split} &\multicolumn{2}{c}{10 vs. 10}\\
\hline
NVD 			&0.8176	&JSD 			&0.5570\\
JSD 			&0.8164	&RNOD 			&0.5272\\
RNOD 			&0.8132	&NVD 			&0.5237\\
RNOD2 			&0.8132	&RNADW 		&0.5198\\
RNADW 			&0.8115	&RNOD2 		&0.5179\\
RNADW2 		&0.8110	&RNADW2 		&0.5165\\
RNSS 			&0.8108	&DNKT\_JSD 	&0.5136\\
DNKT\_JSD 		&0.7814	&RNSS 			&0.5127\\
NMD 			&0.7677	&NMD 			&0.4892\\
DNKT\_RNOD 	&0.7659	&DNKT\_RNOD 	&0.4787\\
DNKT\_NMD 		&0.7490	&DNKT\_NMD 	&0.4737\\
DNKT 			&0.7149	&DNKT 			&0.4288\\
\bottomrule
\end{tabular}
\end{small}
\end{table}




\clearpage

\section{Conclusions}\label{s:conclusions}

As a 
follow-up study of the work reported in \citet{sakai21acl,sakai21cikmlq},
we considered variants of RNOD,
as well as DNKT which is designed to pay attention to the
Priority Preserving Property rather than to the absolute 
probability in each class.
We examined three nominal quantification measures (NVD, RNSS, and JSD),
five ordinal quantification measures
(NMD, RNOD, RNOD2, RNADW, RNADW2), and
DNKT as well as three combined measures,
in terms of system ranking agreement
and system ranking consistency.

From our system ranking agreement results,
we found that, not surprisingly, DNKT is the clear outlier measure
in our suite of measures.
Moreover, we found that 
NMD ranks system differently compared to other measures.
The latter finding 
generalises the results of \citet{sakai21acl,sakai21cikmlq},
which showed that the property of RNOD lies
somewhere between NMD and the nominal quantification measures.

From our system ranking consistency results,
we found that RNOD is the overall winner among the 
ordinal quantification measures:
that is, RNOD outperforms NMD and the variants of RNOD.
Hence, the original design choice of RNOD is reasonable in this respect.
More specifically, assuming equidistance (Eq.~\ref{eq:equidistance})
and considering only classes with a non-zero gold probability (Eq.~\ref{eq:OD})
when computing the Order-aware Divergence are actually beneficial to some degree.
On the other hand, DNKT,
which only cares about whether the classes are prioritised 
as defined in the gold distribution suffers in terms 
of the robustness of the ranking to the choice of test data,
relative to the quantification measures discussed in this study.
Hence, if the PPP is considered important and DNKT seems appropriate for the task,
one should consider this statistical instability into account
at the sample size design stage~\cite{sakai16irj,sakai18book},
i.e., when deciding on the number of test cases for system evaluation.
Recall also that DNKT is not useful if the gold distribution is uniform (See Section~\ref{ss:DNKT}).

\citet{sakai22arxiv-groupfairness} have used JSD, NMD, and RNOD
as a component of their group-fair ranking evaluation measures.
Our future work includes examining the relationship between
ordinal quantification task evaluation 
and the evaluation of other tasks such as group-fair ranking
that involve these divergence measures.

\bibliographystyle{ACM-Reference-Format}
\bibliography{arxiv2022divergence}

\end{document}